\documentclass[apj]{emulateapj}

\usepackage{amsmath}
\usepackage{apjfonts}

\begin{document}

\title{Three-Dimensional Relativistic MHD Simulations of the
  Kelvin-Helmholtz Instability: Magnetic Field Amplification by a
  Turbulent Dynamo}

\author{Weiqun Zhang\altaffilmark{1}, Andrew MacFadyen\altaffilmark{1}
and Peng Wang\altaffilmark{2}}

\altaffiltext{1}{Center for Cosmology and Particle Physics, Physics
  Department, New York University, New York, NY 10003}
\altaffiltext{2}{Kavli Institute for Particle Astrophysics and
  Cosmology, Stanford Linear Accelerator Center and Stanford Physics
  Department, Menlo Park, CA 94025}

\begin{abstract}

  Magnetic field strengths inferred for relativistic outflows
  including gamma-ray bursts (GRB) and active galactic nuclei (AGN)
  are larger than naively expected by orders of magnitude.  We present
  three-dimensional relativistic magnetohydrodynamics (MHD)
  simulations demonstrating amplification and saturation of magnetic
  field by a macroscopic turbulent dynamo triggered by the
  Kelvin-Helmholtz shear instability.  We find rapid growth of
  electromagnetic energy due to the stretching and folding of field
  lines in the turbulent velocity field resulting from non-linear
  development of the instability.  Using conditions relevant for GRB
  internal shocks and late phases of GRB afterglow, we obtain
  amplification of the electromagnetic energy fraction to $\epsilon_B
  \sim 5 \times 10^{-3}$.  This value decays slowly after the shear is
  dissipated and appears to be largely independent of the initial
  field strength.  The conditions required for operation of the dynamo
  are the presence of velocity shear and some seed magnetization both
  of which are expected to be commonplace.  We also find that the
  turbulent kinetic energy spectrum for the case studied obeys
  Kolmogorov's $5/3$ law and that the electromagnetic energy spectrum
  is essentially flat with the bulk of the electromagnetic energy at
  small scales.

\end{abstract}


\section{Introduction}
\label{sec:intro}

Strong magnetic fields are required in GRB and AGN outflows to enable
sufficient production of non-thermal radiation.  Synchrotron models
for GRB afterglow emission require magnetic field strengths many
orders of magnitude larger than expected in smoothly shock compressed
circum-burst medium.  For the prompt GRB emission, magnetic field may
be advected from the central engine to the emission region, but the
degree of advected magnetization is uncertain and may be small.  The
source of sufficiently strong magnetic field responsible for
non-thermal GRB emission has thus remained a mystery, and a similar
situation exists for AGN.  Possible mechanisms capable of generating
magnetic fields in relativistic collisionless shocks are plasma
instabilities, such as the Weibel two-stream instability
\citep{gruzinov99, medvedev99}.  Recent plasma simulations using the
particle-in-cell (PIC) method \citep{buneman} have demonstrated
magnetic field generation in relativistic collisionless shocks
\citep{nishikawa03, spitkovsky08}.  However, the size of the
simulation boxes is orders of magnitude smaller than the GRB and AGN
emitting regions. It remains unclear whether fields generated on
scales of tens of plasma skin depths, as in the current PIC
simulations, will persist at sufficient strength, or at all, on
radiation emission scales.  In addition, the recent plasma simulations
\citep{spitkovsky08} are 
two-dimensional hence missing intrinsically three-dimensional effects
which may be necessary for dynamo operation \citep{gruzinov08}.  Thus
it is far from clear that magnetized regions with size scales and
magnitudes inferred for AGN and GRBs can be generated by plasma
instabilities alone.

Another possibility is that weak magnetic field is greatly amplified
by a magnetohydrodynamic (MHD) turbulent dynamo \citep{kazantsev68,
  kulsrud92, boldyrev04, gruzinov08}.  In our scenario the turbulence
is driven by non-linear development of the Kelvin-Helmholtz
instability present due to velocity shear.  Shear is naturally
expected in realistic explosive outflows due to intermittency and
inhomogeneity in the outflow itself and in the surrounding medium.
Collisions in GRB internal shocks, for example, likely occur between
blobs of differing size and shape resulting in oblique shocks and
vorticity generation in the post-shock flow \citep{goodman08}.  In
jets, shear is intrinsic between a fast jet core and the surrounding
``cocoon'' which is often turbulent due to Kelvin-Helmholtz
instability.  Shocks launched into a clumpy medium, as expected, for
example, in the regions surrounding massive stars, will lead to
significant vorticity generation and velocity shear during clump
shocking \citep{goodman08} or from ion streaming \citep{couch}.  In
these cases the Kelvin-Helmholtz instability is likely to generate
turbulence from the shear flow.  These flows are highly ionized and as
such are expected to contain some magnetic field however weak.

In this letter, we present high-resolution, three-dimensional
numerical simulations of relativistic MHD demonstrating amplification
of weak magnetic fields by Kelvin-Helmholtz instability induced
turbulence.  This process is demonstrated to be viable as the origin
of strong magnetic fields present behind relativistic collisionless
shocks of astrophysical jets in AGN and GRBs.  Our simulations also
allow us to study the basic properties of turbulence in the largely
unexplored relativistic MHD case.  In \S~\ref{sec:init} we describe
the initial setup of our simulations.  In \S~\ref{sec:res} we present
the simulation results.  We further discuss the results in
\S~\ref{sec:dis}.

\begin{figure*}
\plotone{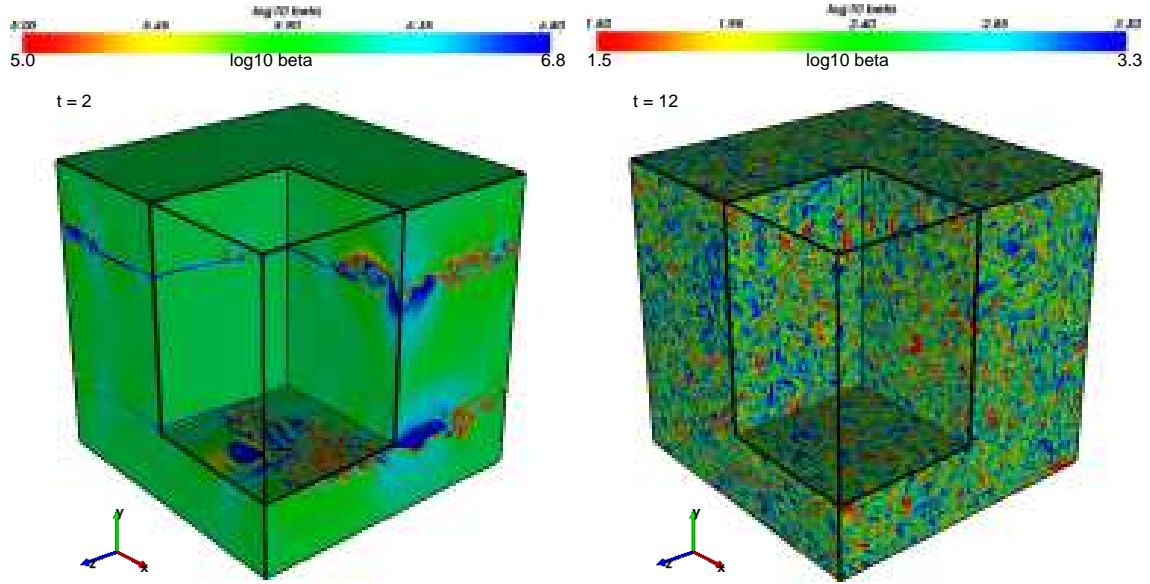}
\caption
{
Plasma $\beta$, the ratio of gas pressure to the magnetic
  pressure, for Model A-1024.  Results at $t=2$ and $t=12$
  are shown in the left and right panel, respectively.  $1024^3$
  numerical cells are used in this calculation for which the numerical box size and light
  crossing time is $1$.  At $t=0$, the speed of the gas is set to $\mathrm{v}_x=0.5$ in the top 
and bottom quarters of the box and $\mathrm{v}_x=-0.5$ in the middle half of the box, with small perturbations added.
  Initially, the box is filled with a weak magnetic field, which
  corresponds to $\beta = 2 \times 10^6$.  By $t=2$, vortices due to the Kelvin-Helmholtz 
  instability are clearly visible.  The instability
  grows and the magnetic field is amplified by stretching and folding of field lines.  By $t=12$, the gas has
  become fully turbulent due to the instability and the plasma beta is highly variable with values as low as $\sim 10$ in some regions.
  \label{fig:beta}}
\end{figure*}

\section{MHD Equations and Initial Setup}
\label{sec:init}

The system of special relativistic MHD (SRMHD) is governed by the
equations of mass conservation, energy-momentum conservation and
magnetic induction, respectively given by
\begin{eqnarray}
\nabla_\nu\rho u^\nu = 0, \\
\nabla_{\nu}T^{\mu\nu} = 0, \\
\frac{\partial}{\partial{t}}\vec{B} = \vec{\nabla} \times
(\vec{v} \times \vec{B}).
\end{eqnarray}
Here $\rho$ is the mass density in the local rest frame, $u^\nu$ is the
four-velocity, $T^{\mu\nu}$ is the energy-momentum tensor, $\vec{B}$
is the magnetic field three-vector in the lab frame, and $\vec{v}$ is
the three-velocity.  The energy-momentum tensor, which includes both
the fluid and electromagnetic parts is given by,
\begin{equation}
  T^{\mu\nu} = \rho h u^\mu u^\nu + P g^{\mu\nu} + (u^\mu u^\nu +
\frac{1}{2} g^{\mu\nu}) b^2 - b^{\mu}b^{\nu},
\end{equation} 
where $P$ is the pressure, $g^{\mu\nu}$ is the metric, $b^\mu$ is the
four-vector of the magnetic field measured in the local rest frame,
and $h \equiv 1 + e + P/\rho $ is the specific enthalpy here $e$ is
the specific internal energy.  Units in which the speed of light is
set to $c = 1$ are used, and the magnetic field is redefined to absorb
a factor of $1/\sqrt{4\pi}$.  The system is closed by an equation of
state.  Furthermore, there is the solenoidal constraint $\vec{\nabla}
\cdot \vec{B} = 0$.

A mildly relativistic and weakly magnetized shear flow with zero net
momentum was set up as the initial condition.   A three-dimensional
grid $(x,y,z)$ in Cartesian coordinates was used with periodic
boundary conditions applied in all directions.  The dimensionless size
of the box was $1$ in each direction and the center of the box was at
$(0,0,0)$.  Initially the numerical box was filled with
relativistically hot gas of constant pressure of $P_0 = 1$ and
constant density of $\rho_0 = 1$. The equation of state was assumed to
be an ideal gas with gamma-law, $P = (\gamma-1) \rho e$, with a
constant adiabatic index $\gamma = 4/3$.  
A shearing flow initial condition was
imposed with the velocity in the top quarter ($y>0.25$) and bottom
quarter ($y<-0.25$) of the box in the positive $x$-direction at a
speed of $v_x = 0.5$ and in the negative $x$-direction at a speed of
$v_x = -0.5$ in the middle half ($-0.25<y<0.25$) of the box. This
corresponds to a shear flow with relative Lorentz factor of
$2.29$. The system had an initially weak magnetic field of $B_x =
10^{-3}$ and $10^{-2}$ for Models A and E, respectively.  Small
velocity perturbations were added of the form,
\begin{equation}
v_{i1} = a_p \cos{(\mathbf{k} \cdot \mathbf{r} + \phi_i)},
\end{equation}
where $i$ denotes $x$-, $y$- and $z$-direction, $\mathbf{r}$ is the
position and $\phi_i$ is a random number in the range of $[0,2\pi)$.
The amplitude of the perturbation is set to $a_p = 0.01$.  The wave
vector $\mathbf{k}$ is given by $\mathbf{k} = 2\pi (\cos{\theta_p}
\mathbf{e}_x + \sin{\theta_p} \mathbf{e}_y)$, here $\theta_p$ is set
to $0.05$.  The initial perturbation is roughly a single mode.  The
relativistic sound speed in the initial models is $c_s = 0.516$ and
the flow speed is $0.5$ (with perturbations of $\leq 0.01$).  Thus the
initial shear flow is transonic.  Also the initial conditions of the
shear flow correspond to $\beta = 1.2 \times 10^6$ and $1.2 \times
10^4$ for Models~A and E, respectively.  Here the plasma beta is
$\beta = P/P_{m}$, where $P$ is gas pressure and $P_m$ is magnetic
pressure.  In the absence of magnetic field, Kelvin-Helmholtz
instability develops no matter how small or large the velocity shear
is, whereas strong magnetic field can suppress the instability.  This
is true in 3D even for supersonic flow speeds because there are always
directions along which the shear flow component is subsonic \citep{landau}.
Because the initial field in our models is small and dynamically
unimportant, Kelvin-Helmholtz instability develops in our 3D
simulations.

\section{Results}
\label{sec:res}

The SRMHD  simulations in this letter were performed
with the general relativistic MHD code Nvwa (Zhang \& Wang in prep).
The three-dimensional numerical simulations were run with various
resolutions: $512^3$, $768^3$ and $1024^3$ numerical cells.  We use
A-n and E-n to denote Models A and E, respectively, where n stands for
the number of numerical cells in each dimension.  For example,
Model~A-1024 stands for Model~A with a resolution of $1024^3$
numerical cells.

\subsection{Amplification of Magnetic Fields}
\label{sec:amp}

\begin{figure}
\plotone{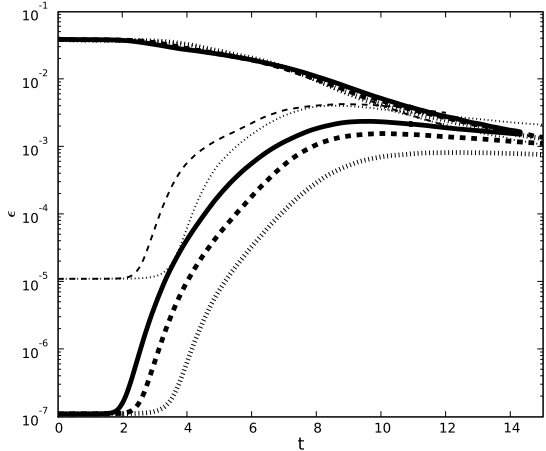}
\caption{Ratio of electromagnetic and kinetic energy to total energy
  as a function of time for Models A and E.  
  The lines for kinetic energy start at $\epsilon \sim 4 \times
  10^{-2}$ at $t=0$ for all models, whereas the lines for
  electromagnetic energy start at $\epsilon \sim 10^{-7}$ and
  $10^{-5}$ for Models~A and E respectively.    
  Different lines are for
  different resolutions: $1024^3$ cells ({\it solid line}), $768^3$
  cells ({\it dashed lines}), and $512^3$ cells ({\it dotted lines}).
  Models A and E are shown in thick and thin lines, respectively.  The
  ratio of electromagnetic energy to total energy ($\epsilon_B$) grows
  due to growth of the Kelvin-Helmholtz instability and ensuing turbulence.  The electromagnetic energy
  saturates at $t \sim 9$ depending upon the model and resolution.  The
  kinetic energy decays over time, and it becomes comparable to
  electromagnetic energy at $t \sim 14$ for Model~A-1024.
  \label{fig:eb}}
\end{figure}

In all models the initial discontinuity in velocity triggered
Kelvin-Helmholtz instability.  Two snapshots of Model~A-1024 are shown
in Fig.~\ref{fig:beta}.  Kelvin-Helmholtz vortices start to appear on
the plane of the velocity discontinuity after $\sim 2$ light crossing
times of the numerical box due to the roll-up of the shear flow.  The
magnetic field at the shear layer is being folded and twisted by
the vortical fluid motion.  The strength of the
magnetic field starts to increase rapidly, growing exponentially as
shown in Fig.~\ref{fig:eb}.  
As the instability evolves,
more and more gas is turned into a state of turbulence.  At about $t
\sim 7$, the whole box has become fully turbulent, and the
electromagnetic energy starts to saturate.  The turbulence eventually
decays because there is no bulk shear flow left to further drive the
turbulence.  Fig.~\ref{fig:eb} shows the evolution of the ratio of the
electromagnetic energy to total energy for Models~A and E at various
levels of resolution, here the total energy does not include rest mass
energy.  Also shown in Fig.~\ref{fig:eb} is the ratio of the kinetic
energy\footnote{The kinetic energy density is defined as $\rho \Gamma
  (\Gamma -1)$, where $\rho$ is mass density in the fluid rest frame and
  $\Gamma$ is the Lorentz factor in the laboratory frame.} to total
energy.  Note that the kinetic energy decays faster than the
electromagnetic energy.  For Model~A-1024, the electromagnetic energy
becomes comparable to kinetic energy at $t \sim 14$.

The spatial structure of plasma beta for Model~A-1024 is shown in
Fig~\ref{fig:beta} at two times.  The structure is clumpy and
filamentary as expected for magnetic fields amplified by a turbulent
dynamo.  At the stage of full turbulence, the average ratio of the
electromagnetic energy to total energy is $\epsilon_B \sim 2 \times
10^{-3}$ for Model~A-1024, whereas maximum values of
$\epsilon_{B,\mathrm{max}} \sim 0.2$ exist throughout the box.  Thus
magnetic field is very strong in local patches of the turbulence.  
These strongly magnetized clumps are preferably elongated along the
magnetic field direction due to the intrinsic anisotropy
of MHD turbulence \citep{gs95}.   
It should be noted that there is no overall magnetic flux except for a
very low magnetic flux in $x$-direction present in the initial
conditions.  The strength of the magnetic field is a result of folding
and twisting of existing field lines by turbulent motions.  This process
increases magnetic energy density without increasing total magnetic
flux.

To study the effects of numerical resolution and initial magnetic
field strength on the level of magnetization that can be reached, we
have run a series of calculations, Models~A-768, A-512, E-768 and
E-512 for comparison (Fig.~\ref{fig:eb}).  As expected, the onset of
the instability is delayed and the saturated level of magnetization is
lower for lower resolution runs due to their higher numerical
diffusivity.  The maximal overall ratio of the electromagnetic energy
to total energy reaches $8.2 \times 10^{-4}$, $1.6 \times 10^{-3}$ and
$2.4 \times 10^{-3}$ for Models~A-512, A-768 and A-1024, respectively.
For Model~E, which started with higher initial magnetic field, the
final magnetization is higher than that of Model~A as we also expected
because less stretching and folding is required for reaching the same
magnetic field strength.  The maximal overall ratio of the
electromagnetic energy to total energy reaches $4.1 \times 10^{-3}$
and $4.5 \times 10^{-3}$ for Models~E-512 and E-768, respectively.
But it should be noted that the two orders of magnitude difference in
the initial fields causes only a factor of $<3$ in the final ratio of
the electromagnetic energy to total energy.  These results suggest
that the ``true'' answer is $\epsilon_B \sim 5 \times 10^{-3}$ for the
chosen conditions.

\subsection{MHD Turbulence}
\label{sec:turb}

\begin{figure}
\plotone{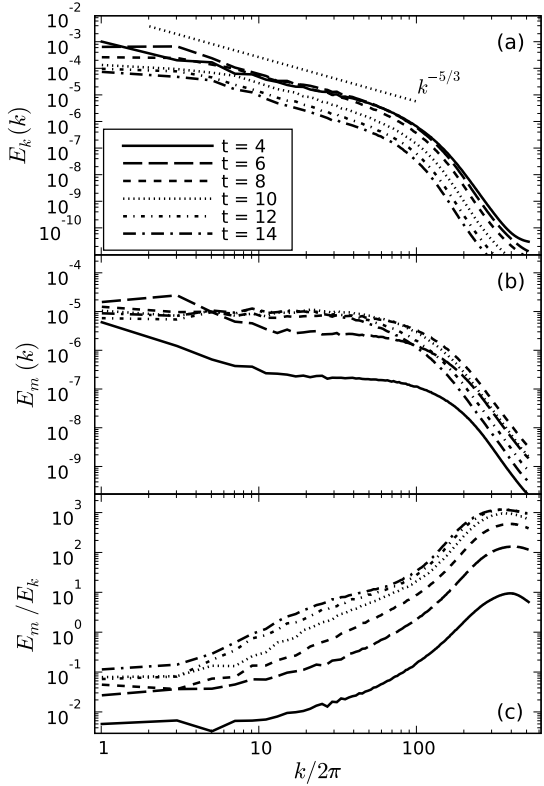}
\caption{Spherically-integrated spectrum of (a) kinetic energy ({\it
    top}) and (b) electromagnetic energy ({\it middle}), and (c) ratio
  of electromagnetic spectrum to kinetic energy spectrum ({\it bottom})
  for Model~A-1024.
  Different lines are for different times: $t=4$ ({\it solid lines}),
  $t=6$ ({\it long dashed lines}), $t=8$ ({\it short dashed lines}),
  $t=10$ ({\it dotted lines}), $t=12$ ({\it short dash-dotted lines}),
  and $t=14$ ({\it long dash-dotted lines}).
  Also shown in (a) is a line representing the power law $E(k) \sim
  k^{-5/3}$ for comparison.   
  \label{fig:pk}}
\end{figure}

We now analyze the properties of decaying mildly relativistic
adiabatic MHD turbulence using the data of Model~A-1024.  The initial
flow is mildly relativistic and transonic with shear velocity of $0.5$
corresponding to a Lorentz factor of $\Gamma_0 = 1.15$.  During the
evolution, the velocities of plasma elements decrease due to
dissipation into heat and the work done to stretch and fold magnetic
field lines.  The typical velocities in Model~A-1024 at $t = 4$, $8$,
$14$, are about $0.4$, $0.3$ and $0.1$, with a standard deviation of
$0.1$, $0.1$ and $0.05$ respectively.  Hence, the MHD turbulence in our
simulations is mostly subsonic except for at the very beginning.

We define the kinetic energy spectrum $E_k(k)$ as 
\begin{equation}
\int{E_k(k) dk} = \langle\rho \Gamma (\Gamma-1)\rangle,
\end{equation}
where $\Gamma$ is the Lorentz factor, and the electromagnetic energy
spectrum $E_m(k)$ as
\begin{equation}
 \int{E_m(k)dk} = \langle T^{00}_{\mathrm{EM}}\rangle,
\end{equation}
where $T^{00}_{\mathrm{EM}}$ is the $00$-component of the
energy-momentum tensor for electromagnetic field.

It is clearly shown in Fig.~\ref{fig:pk} that the kinetic energy
spectra follow the Kolmogorov's $5/3$ law in the inertial range and
fall off at $k/2\pi \sim100$.  The spectrum is slightly flatter than
$k^{-5/3}$ before the saturation of the overall electromagnetic
magnetic energy at $t \sim 9$, whereas it is slightly steeper than
$k^{-5/3}$ after the saturation.  The kinetic energy decreases over
time due to viscous dissipation.  

The electromagnetic energy spectrum is also shown in
Fig.~\ref{fig:pk}.  Note that the electromagnetic energy spectrum
increases over time until it saturates.  But the overall shapes of the
spectrum at $t = 4$ and $6$ are very similar.  This is a typical
behavior of the so-called small-scale dynamo \citep{kazantsev68}.  The
electromagnetic energy saturates at small scales ($k/2\pi>100$) and
large scales ($k/2\pi<5$) first, then at intermediate scales
($5<k/2\pi<100$) later.  There appears to be both a forward cascade of
electromagnetic energy from large scales to intermediate scales and an
inverse cascade from small scales to intermediate scales.  It is
striking that the electromagnetic energy spectra are flat and evolve
very slowly after $t = 8$, the time when the electromagnetic energy
saturates.  It is not surprising that $t = 8$ is also the time when
the whole simulation box is filled with turbulence.  A flat magnetic
energy spectrum is commonly seen in turbulent dynamo simulations
\citep[e.g.,][]{brandenburg01, schekochihin04}.  The shape of
electromagnetic spectra clearly indicate that the bulk of the
electromagnetic energy is at small scales.

Fig.~\ref{fig:pk} shows that the ratio of electromagnetic to kinetic
energy spectrum increases monotonically on almost all scales as the
instability and turbulence develop.  The electromagnetic energy
dominates kinetic energy at small scales, whereas the kinetic energy
dominates at large scales.  The equipartition point moves toward
larger scales during the evolution.  Both electromagnetic and kinetic
energy are decreasing after $t \sim 9$ (Fig.~\ref{fig:eb}). However,
the kinetic energy decreases much faster than the electromagnetic
energy.  This means that the electromagnetic resistive dissipation is
less efficient than the viscous dissipation of kinetic energy.
Therefore the magnetic fields amplified by the turbulent dynamo can
persist for a very long time.

\section{Discussion}
\label{sec:dis}

In this letter, we have presented a series of relativistic MHD
simulations with an initial density of $\rho_0 = 1$, pressure of $P_0
= 1$, and shear velocity of $\mathrm{v}_0 = 0.5$.  The simulations are
representative of conditions in which a turbulent dynamo operates
behind a relativistic astrophysical shock.  The initial conditions for
models A and E presented here are relevant for GRB internal shocks,
the late stage of GRB afterglows, trans-relativistic explosions of
1998-bw like supernovae, {\emph{aka}} ``hypernovae'' \citep{SN1998bw},
and shock breakout from Type Ibc supernova SN2008 \citep{sn2008D}.
Our calculations predict the value of the ratio of magnetic energy to
total energy, $\epsilon_B = 5 \times 10^{-3}$, which is typically
treated as a free parameter in radiation modelling.  Further work is
under way to cover more parameter space relevant to additional stages
of GRB and AGN outflows, and other astrophysical relativistic outflows
such as those from magnetars \citep{gelfand} and microquasars \citep{mirabel}.  These results and a
comprehensive study of relativistic turbulence now underway will be
presented in future publications.

Our simulations show that the electromagnetic energy structure is very
clumpy.  This turbulent shearing gas with moving magnetic clouds is likely a
site for Fermi acceleration of charged particles, and thus a source of
high energy cosmic rays.  We are currently calculating charged particle
acceleration in model A (Zrake et al., in prep).  It is possible that a
self-consistent model of GRB afterglows, which includes the
acceleration of nonthermal electrons and the amplification of magnetic
fields, can be built upon the processes considered in this letter.
The spatial variability and large fluctuations of the turbulent
magnetic field also have implications for non-thermal radiation which
may have observable consequences.  Synchrotron and inverse Compton
spectra from the turbulent fields are being calculated and will be
presented in a future paper (van Velzen et al., in prep).

The turbulent dynamo studied in this letter may also take place in
other astrophysical environments.  It is still a mystery how the
cosmic magnetic fields in galaxies and clusters of galaxies got
started \citep[see][for a review]{kulsrud08}.  One possibility is that
the cosmic magnetic fields are generated by turbulence during
hierarchical structure formation \citep{pudritz89, kulsrud97}.
Another possibility is that the strong magnetic field in the jets of
AGN could spread to the entire universe \citep{rees68, daly90}.  The
Kelvin-Helmholtz induced turbulent dynamo we have considered can be at
work in both cases to provide a seed field for the mean-field dynamo
\citep[][and references therein]{kulsrud08}.

\acknowledgments 

We are greatly indebted to Andrei Gruzinov for many stimulating
discussions.  We would also like to thank Yosi Gelfand and Martin
Pessah for useful discussions. This research utilized resources at the
New York Center for Computational Sciences at Stony Brook
University/Brookhaven National Laboratory which is supported by the
U.S. Department of Energy under Contract No. DE-AC02-98CH10886 and by
the State of New York and the CCNI, supported by the New York State
Foundation for Science, Technology and Innovation (NYSTAR).


\begin{thebibliography}{}

\bibitem[Boldyrev \& Cattaneo(2004)]{boldyrev04} 
Boldyrev, S., \& Cattaneo, F.\ 2004, Physical Review Letters, 92, 144501 

\bibitem[Brandenburg(2001)]{brandenburg01} 
Brandenburg, A.\ 2001, \apj, 550, 824 

\bibitem[Buneman(1993)]{buneman}
Buneman, O. 1993, in Computer Space Plasma Physics: Simulation
Techniques and Software, ed. H. Matsumoto \& Y. Omura (Tokyo: Terra
Scientific Publ. Co.), 67

\bibitem[Couch, Milosavljevi{\'c} \& Nakar(2008)]{couch} Couch, S.~M., 
Milosavljevi{\'c}, M., \& Nakar, E.\ 2008, \apj, 688, 462

\bibitem[Daly \& Loeb(1990)]{daly90} 
Daly, R.~A., \& Loeb, A.\ 1990, \apj, 364, 451 

\bibitem[Gelfand et al.(2005)]{gelfand} Gelfand, J.~D., et al.\ 
2005, \apjl, 634, L89 

\bibitem[Goldreich \& Sridhar(1995)]{gs95}
Goldreich, P., \& Sridhar, S.\ 1995, \apj, 438, 763 

\bibitem[Goodman \& MacFadyen(2008)]{goodman08}
Goodman, J., \& MacFadyen, A.\ 2008, Journal of Fluid Mechanics, 604, 325

\bibitem[Gruzinov \& Waxman(1999)]{gruzinov99} 
Gruzinov, A., \& Waxman, E.\ 1999, \apj, 511, 852 

\bibitem[Gruzinov(2008)]{gruzinov08} 
Gruzinov, A.\ 2008, arXiv:0803.1182 

\bibitem[Kazantsev(1968)]{kazantsev68} 
Kazantsev, A.~P.\ 1968, Soviet Journal of Experimental and Theoretical
Physics, 26, 1031  

\bibitem[Kulkarni et al.(1998)]{SN1998bw} Kulkarni, S.~R., et 
al.\ 1998, \nat, 395, 663 

\bibitem[Kulsrud \& Anderson(1992)]{kulsrud92} 
Kulsrud, R.~M., \& Anderson, S.~W.\ 1992, \apj, 396, 606 

\bibitem[Kulsrud et al.(1997)]{kulsrud97} 
Kulsrud, R.~M., Cen, R., Ostriker, J.~P., \& Ryu, D.\ 1997, \apj, 480, 481 

\bibitem[Kulsrud \& Zweibel(2008)]{kulsrud08} 
Kulsrud, R.~M., \& Zweibel, E.~G.\ 2008, Reports on Progress in Physics, 71, 046901 

\bibitem[Landau \& Lifshitz(1959)]{landau} 
Landau, L.~D., \& Lifshitz, E.~M.\ 1959, Fluid Mechanics (Oxford:
Pergamon)

\bibitem[Medvedev \& Loeb(1999)]{medvedev99} 
Medvedev, M.~V., \& Loeb, A.\ 1999, \apj, 526, 697 

\bibitem[Mirabel et al.(1992)]{mirabel} Mirabel, I.~F., 
Rodriguez, L.~F., Cordier, B., Paul, J., 
\& Lebrun, F.\ 1992, \nat, 358, 215 


\bibitem[Nishikawa et al.(2003)]{nishikawa03} 
Nishikawa, K.-I., Hardee, P., Richardson, G., Preece, R., Sol, H., 
\& Fishman, G.~J.\ 2003, \apj, 595, 555 

\bibitem[Pudritz \& Silk(1989)]{pudritz89} 
Pudritz, R.~E., \& Silk, J.\ 1989, \apj, 342, 650 

\bibitem[Rees \& Setti(1968)]{rees68} 
Rees, M.~J., \& Setti, G.\ 1968, \nat, 219, 127 

\bibitem[Schekochihin et al.(2004)]{schekochihin04} 
Schekochihin, A.~A., Cowley, S.~C., Taylor, S.~F., Maron, J.~L., 
\& McWilliams, J.~C.\ 2004, \apj, 612, 276 

\bibitem[Soderberg et al.(2008)]{sn2008D} Soderberg, A.~M., et 
al.\ 2008, \nat, 453, 469 

\bibitem[Spitkovsky(2008)]{spitkovsky08} 
Spitkovsky, A.\ 2008, \apjl, 682, L5 

\end{thebibliography}
\end{document}